\documentclass[onecolumn,a3paper,10pt]{article}
\usepackage{graphicx}
\usepackage{sidecap}% Include figure files
\usepackage{bm}% bold math
\usepackage{color}
\usepackage{ulem}
\usepackage[colorlinks,linkcolor=blue, urlcolor=blue, anchorcolor=blue, citecolor=blue]{hyperref}
\usepackage{amsmath,amsfonts,amssymb}
\usepackage{booktabs}
\usepackage{float}
\usepackage{mathrsfs}
\usepackage{cite}

\title{Shortcuts to adiabaticity in five-level systems using counter-diabatic driving and time-rescaling optimization}

\author{Jiahui Zhang\footnote{Corresponding author:  Jia-Hui Zhang, E-mail: jhzhang\_phys@nwnu.edu.cn}, Wenyuan Wang and Fuquan Dou}

\begin{document}

\maketitle
\noindent \small{College of Physics and Electronic Engineering, Northwest Normal University, Lanzhou 730070, China}

\begin{abstract}
Shortcuts to adiabaticity (STA) is a common protocol to realize high-fidelity and robust quantum control in various quantum systems. To date, STA has been widely applied in two- and three-level systems, whereas designing feasible strategies to achieve perfect quantum state engineering in multi-level systems still remains a challenging task. Here, we propose to use counterdiabatic (CD) driving and time-rescaling (TR) methods to construct multi-state stimulated Raman shortcut-to-adiabatic passage protocols for realizing robust and fast population transfer in chainwise-connected five-state systems. The first protocol is implemented by initially reducing the original five-state system to an equivalent two-state counterpart, and then designing the corresponding driving field by combining CD driving and unitary transformation. Further, we introduce the TR method to optimize the first protocol and thus offers an alternative solution. Numerical calculations show that both protocols can achieve complete population transfer and effectively suppress transient populations of all intermediate states. Compared with the first protocol, the optimized second exhibits better performance within a shorter evolution time.
\end{abstract}
%\noindent \textbf{Corresponding author}:  Jia-Hui Zhang, E-mail: jhzhang\_phys@nwnu.edu.cn

As a mainstream variant of stimulated Raman adiabatic passage (STIRAP), multi-state STIRAP (hereafter abbreviated as ``M-STIRAP") is a pivotal technique to achieve high-fidelity population transfer between the two end states of a chainwise-connected multi-level system~\cite{10.1063/1.4916903, Vitanov2017}.
It has found extensive applications in physics, chemistry, and beyond~\cite{Bergmann2019}. However, the successful implementation of STIRAP-like process relies on adiabatic evolution, which requires long operation times and may deteriorate the practical control performance. 
Thus, it is natural to search for novel methods which are fast and robust to improve or replace the adiabatic passage techniques~\cite{V2025, McCord2025}.
%This is problematic as the system will therefore also have a long time to interact with an environment leading to losses or decoherence.
%To make the STIRAP approach more practical, it is necessary to make adiabatic operations as fast and robust as possible.

In order to make adiabatic operations as fast and robust as possible, a method dubbed ``shortcuts to adiabaticity (STA)'' has been proposed~\cite{Chen2010, Gu2019, delCampo2019, Hatomura2024}. STA is the collective term for a family of approaches designed to speed up adiabatic passage. One particularly successful method for constructing such fast routes is counterdiabatic (CD) driving~\cite{Demirplak2003, Demirplak2008, Berry2009, Masuda2015, Du2016}. Its basic idea lies in introducing additional CD driving fields to remove unwanted nonadiabatic effects, thereby greatly enhancing the fidelity of adiabatic passage. However, the presence of extra coupling terms poses challenges to practical implementations.
To overcome this obstacle, a series of STA protocols free of extra couplings have been put forward, such as multiple Schr\"odinger dynamics~\cite{PhysRevLett.109.100403, PhysRevA.93.052324}, dressed-state~\cite{PhysRevLett.116.230503, Zhou2017, Wu:17}, invariant-based inverse engineering~\cite{PhysRevLett.104.063002, PhysRevA.86.033405, Ho:15, Benseny2017, 10.1063/5.0183063} and stimulated Raman shortcut-to-adiabatic passage (STIRSAP)~\cite{PhysRevA.94.063411, Du2016}. Despite providing different shortcuts, they are closely related due to their similar underlying physical mechanisms~\cite{WU201740}. Among them, STIRSAP has attracted attention due to its ability to suppress the transient populations of the intermediate state. In most cases, this state is inherently lossy~\cite{PhysRevA.56.1463, Li:17, Zhang2021, 10.1098/rsta.2021.0283}, and any population residing in it will induce a loss of coherence or unwanted decay. This protocol was initially proposed by Li $et$ $al$.~\cite{PhysRevA.94.063411} and later experimentally demonstrated by Du $et$ $al$.~\cite{Du2016}.  Recently, researchers have proposed several optimized solutions for STIRSAP~\cite{Mortensen_2018, Song:21, Messikh_2022, Zheng2022}. In addition, a method
called ``time-rescaling" (TR) has recently been proposed for constructing STA~\cite{PhysRevResearch.2.013133, e23010081, daSilvaAndrade2022, PRXQuantum.5.010322, qc91-5mj2, Montenegro2026}. It enables the generation of the Hamiltonian for fast evolution processes without acquiring the instantaneous eigenstates of the reference protocol.

At present, STA has been widely applied to various two- and three-level quantum systems, while relevant investigations on multi-level systems, which act as vital platforms in quantum field, remain relatively limited~\cite{Vitanov2020, Han2025}. The development of multi-level STA protocols can not only expand the scope of such methods, but also may provide theoretical support for applications based on multilevel systems~\cite{PhysRevA.78.021402, WANG20123498, Huang2022, PhysRevA.91.023802, Amiri_2023, PhysRevA.97.033407, Wang2024, Yuan2024, Zahia2025}.

This work focuses on STA protocols for fast and efficient population transfer in chainwise-connected five-state systems. First, we reduce the original system to an equivalent two-level model, then modify the driving fields via CD driving and unitary transformation, and thereby develop the M-STIRSAP protocol. Second, we utilize the TR technique to optimize the driving fields of M-STIRSAP and obtain its optimized variant M-STIRSAP-opt. The results verify that both protocols work well, while M-STIRSAP-opt features better performance within a shorter timescale. In addition, both protocols can effectively suppress the transient populations on all intermediate levels.

\section{\label{sec:level2}Model and Method}
\subsection{\label{sec:level2.1}The M-STIRSAP}

\begin{figure}
\centering{\includegraphics[width=8cm]{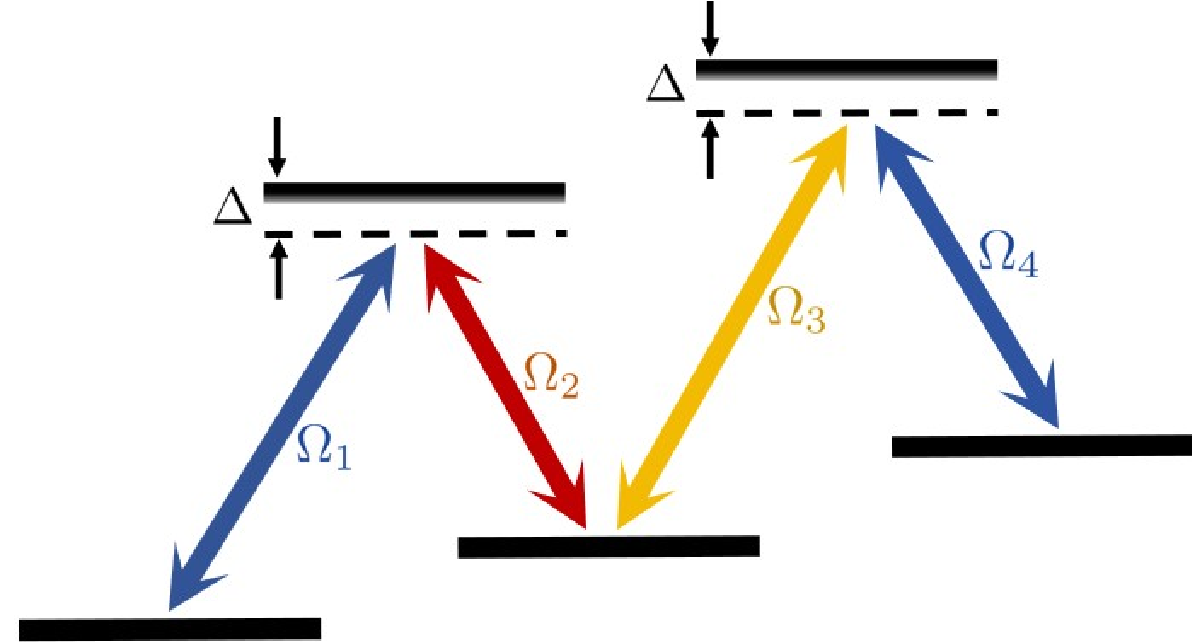}}
\caption{Schematic of an M-type chainwise-connected five-level system.}
\label{fig1}
\end{figure}

We start by considering a chainwise-connected five-level system as shown in Fig.~\ref{fig1}.
The Hamiltonian of the system under the rotating-wave approximation reads $(\hbar\!=\!1)$
\begin{eqnarray}\label{1}
H_5\!=\frac{\rm{1}}{\rm{2}}\!
\left(
\begin{array}{ccccc}
0&\Omega_{\rm{1}}(t)&0&0&0\\
\Omega_{\rm{1}}(t)&2\delta&\Omega_{\rm{2}}(t)&0&0\\
0&\Omega_{\rm{2}}(t)&0&\Omega_{\rm{3}}(t)&0\\
0&0&\Omega_{\rm{3}}(t)&2\delta&\Omega_{\rm{4}}(t)\\
0&0&0&\Omega_{\rm{4}}(t)&0\\
\end{array}
\right).
\end{eqnarray}
where $\Omega_{i}(t) (i\!=\!1, 2, 3, 4)$ denotes the Rabi frequency for the transition between adjacent levels, and $\delta$ denotes single(three)-photon detuning. In our method, we assume that the four Rabi frequencies satisfy the following relation:
\begin{eqnarray}\label{2}
\Omega_{\rm{0}}\!\equiv\!\Omega_{\rm{1}}(t)\!=\!\Omega_{\rm{4}}(t)\!=\!\sqrt{\Omega_{\rm{2}}^2(t)\!+\!\Omega_{\rm{3}}^2(t)}.
\end{eqnarray}
%where $\Omega_{\rm{0}}$ is temporarily introduced to simplify subsequent derivations. 
This relation underpins the subsequent theoretical analysis.

\sloppy
The internal dynamical evolution of the system obeys the time-dependent Schr\"odinger equation (TDSE)
\begin{eqnarray}\label{3}
i\frac{\partial}{\partial t}|\psi(t)\rangle\!=\!H|\psi(t)\rangle,
\end{eqnarray}
where $|\psi(t)\rangle=\left[c_1(t), c_2(t), c_3(t), c_4(t), c_5(t)\right]^{T}$ represents a $5$-dimensional column vector, and $|c_i(t)|^2$ stands for the population of state $|i\rangle$.

We first consider the use of the CD driving method to develop an STA protocol for accelerating the M-STIRAP process. However, the calculation of CD driving requires information on the instantaneous eigenstates of the original multi-level system, a fact that poses difficulties for the implementation in multi-level systems. To remedy this issue, we adopt the adiabatic elimination principle to simplify the above Hamiltonian.
To ensure the validity of AE, we set $\delta\!\gg\!\Omega_{i} (i\!=\!1,2,3,4)$. As a result, we obtain an effective three-level Hamiltonian, which takes the form of
\begin{eqnarray}\label{4}
H_3
=\frac{\rm{1}}{\rm{2}}
\left(
\begin{array}{ccc}
\delta_{\rm{1}}&\omega_{\rm{1}}&0\\
\omega_{\rm{1}}&\delta_{\rm{2}}&\omega_{\rm{2}}\\
0&\omega_{\rm{2}}&\delta_{\rm{3}}\\
\end{array}
\right),
\end{eqnarray}
in which
$\delta_{\rm{0}}\!=\!\delta_{\rm{1}}\!=\!\delta_{\rm{2}}\!=\!\delta_{\rm{3}}\!=\!-{\Omega_{\rm{0}}^2}/{\rm{2}\delta}$, $\omega_{\rm{1}}\!=-\frac{\Omega_{\rm{1}}\Omega_{\rm{2}}}{\rm{2}\delta}, \omega_{\rm{2}}\!=\!-\frac{\Omega_{\rm{3}}\Omega_{\rm{4}}}{\rm{2}\delta}$. By further setting $c_j(t)\!=\!c^{'}_{j}(t)e^{-i\delta_0 t}(j\!=\!1, 3, 5)$, we thus obtain
\begin{eqnarray}\label{5}
H_3^{'}=\frac{\rm{1}}{\rm{2}}
\left(
\begin{array}{ccc}
0&\omega_{\rm{1}}&0\\
\omega_{\rm{1}}&0&\omega_{\rm{2}}\\
0&\omega_{\rm{2}}&0\\
\end{array}
\right),
\end{eqnarray}
where the effective Rabi frequencies are defined as
\begin{eqnarray}\label{6}
\omega_{\rm{1}}\!=\!-\frac{\Omega_{\rm{2}}\sqrt{\Omega_{\rm{2}}^2\!+\!\Omega_{\rm{3}}^2}}{\rm{2}\delta},\nonumber\\
\omega_{\rm{2}}\!=\!-\frac{\Omega_{\rm{3}}\sqrt{\Omega_{\rm{2}}^2\!+\!\Omega_{\rm{3}}^2}}{\rm{2}\delta}.
\end{eqnarray}

Since the resulting Hamiltonian is formally equivalent to the STIRAP Hamiltonian and features the simplest resonant coupling, the system can be further reduced to an effective two-level description,
\begin{eqnarray}\label{7}
H_2=\frac{\rm{1}}{\rm{2}}
\left(
\begin{array}{cc}
-\delta_e&\omega_{e}\\
\omega_{e}&\delta_e\\
\end{array}
\right),
\end{eqnarray}
where the effective Rabi frequency and the detuning are given by
\begin{eqnarray}\label{8}
\omega_e\!=\!\omega_{1}/\rm{2},\nonumber\\
\delta_e\!=\!-\omega_{2}/\rm{2}.
\end{eqnarray}

We now focus on the two-level system~(\ref{7}). By incorporating CD driving~\cite{PhysRevA.94.063411}, we recast the effective two-level Hamiltonian as
\begin{eqnarray}\label{9}
H_2^{'}\!=\!\frac{\rm{1}}{\rm{2}}
\left(
\begin{array}{cc}
-\delta_e&\tilde{\omega}_e^{-i\varphi}\\
\tilde{\omega}_e^{i\varphi}&\delta_e\\
\end{array}
\right),
\end{eqnarray}
where $\tilde{\omega}_e\!=\!\sqrt{\omega^2_{e}\!+\!\omega^2_{cd}},
\omega_{cd}\!=\!(\dot{\omega}_{1}{\omega}_{2}\!-\!\dot{\omega}_{2}{\omega}_{1})/({\omega}^2_{1}\!+\!{\omega}^2_{2})
$, and $\varphi\!=\!\arctan{(\omega_{cd}/\omega_{e})}$. 

To realize counterdiabatic driving without introducing extra couplings, we further perform a unitary transformation and thus obtain,  %$\tilde{H_2}\!=\!U^\dagger H_2^{'}U-i\hbar U^\dagger\dot{U}$,
\begin{eqnarray}\label{10}
\tilde{H_2}\!=\!\frac{\rm{1}}{\rm{2}}
\left(
\begin{array}{cc}
-\tilde{\delta}_e&\tilde{\omega}_e\\
\tilde{\omega}_e&\tilde{\delta}_e\\
\end{array}
\right),
\end{eqnarray}
where
\begin{eqnarray}\label{11}
U\!=\frac{\rm{1}}{\rm{2}}\!
\left(
\begin{array}{cc}
e^{-i\varphi/2}&0\\
0&e^{i\varphi/2}\\
\end{array}
\right),
\end{eqnarray}
and $\tilde{\delta}_e=\delta_e\!+\!\dot{\varphi}$. 

Now we go back to the three-level problem and try to design the corresponding Rabi frequencies.
According to Eq.~(\ref{8}), we may impose $\tilde{\omega}_e\!=\!\tilde{\omega}_{\rm{1}}/\rm{2}$ and $\tilde{\delta}_e\!=\!-\tilde{\omega}_{\rm{2}}/\rm{2}$. 
As a result, the three-level Hamiltonian in Eq.~(\ref{5}) is reformulated as
\begin{eqnarray}\label{12}
\tilde{H_3}=\frac{\rm{1}}{\rm{2}}
\left(
\begin{array}{ccc}
0&\tilde{\omega}_{\rm{1}}&0\\
\tilde{\omega}_{\rm{1}}&0&\tilde{\omega}_{\rm{2}}\\
0&\tilde{\omega}_{\rm{2}}&0\\
\end{array}
\right).
\end{eqnarray}
In principle, the above Hamiltonian enables fast population transfer from $|1\rangle$ to $|5\rangle$. However, such two-photon coupling might be experimentally challenging~\cite{Zhang2025, doi:10.1063/1.2828985, PhysRevA.89.053408}. For this reason, our remaining task is to construct the Rabi frequencies for the five-level system. Like Eqs.~(\ref{6}), we impose
\begin{eqnarray}\label{13}
\tilde{\omega}_{1}\!=\!-\frac{\tilde{\Omega}_2\sqrt{\tilde{\Omega}_{\rm{2}}^2\!+\!\tilde{\Omega}_{\rm{3}}^2}}{2\delta},\nonumber\\
\tilde{\omega}_{2}\!=\!-\frac{\tilde{\Omega}_3\sqrt{\tilde{\Omega}_{\rm{2}}^2\!+\!\tilde{\Omega}_{\rm{3}}^2}}{2\delta}.
\end{eqnarray}
Meanwhile, guided by Eq.~(\ref{2}), we impose
\begin{eqnarray}\label{14}
\tilde{\Omega}_{1, 4}\!=\!\sqrt{\tilde{\Omega}_{2}^2\!+\!\tilde{\Omega}_{3}^2}.
\end{eqnarray}

Solving the above equations yields
\begin{eqnarray}\label{15}
%\begin{aligned}
\tilde{\Omega}_{2}\!=\!\tilde{\omega}_{1}\left[\frac{4\delta^2}
{\tilde{\omega}^2_{1}\!+\!\tilde{\omega}^2_{2}}\right]^{\frac{\rm{1}}{\rm{4}}},\nonumber\\
\tilde{\Omega}_{3}\!=\!\tilde{\omega}_{\rm{2}}\left[\frac{4\delta^2}
{\tilde{\omega}^2_{\rm{1}}\!+\!\tilde{\omega}^2_{\rm{2}}}\right]^\frac{1}{4},\nonumber\\
\tilde{\Omega}_{1, 4}\!=\!\Big\{4\delta^2\left[\tilde{\omega}^2_{\rm{1}}\!+\!\tilde{\omega}^2_{\rm{2}}\right]\Big\}^{\frac{1}{4}}.
%where $\tilde{\omega}_{0}\!=\!\tilde{\omega}^2_{1}\!+\!\tilde{\omega}^2_{2}$.
%\end{aligned}
\end{eqnarray}
\noindent
%The remaining Rabi frequencies can be modified in an analogous manner to that described in Eq.~(\ref{2}), namely,
%$\tilde{\Omega}_{1, 4}\!=\!\sqrt{\tilde{\Omega}_{2}^2\!+\!\tilde{\Omega}_{3}^2}.$
%Accordingly, we can express them as
%\begin{eqnarray}\label{14}
%\tilde{\Omega}_{1, 4}\!=\!\Big\{4\delta^2\left[\tilde{\omega}^2_{\rm{1}}\!+\!\tilde{\omega}^2_{\rm{2}}\right]\Big\}^{\frac{1}{4}}.
%\end{eqnarray}

%\noindent
By substituting the original Rabi frequencies in Hamiltonian~(\ref{1}) with four modified versions and imposing the condition $\delta\!\gg\!\tilde{\Omega}_{j}(j\!=\!1,2,3,4)$, we can recast the five-level Hamiltonian as
\begin{equation}\label{16}
\tilde{H}_5=\frac{1}{2}
\left(
\begin{array}{ccccc}
0&\tilde{\Omega}_{\rm{1}}&0&0&0\\
\tilde{\Omega}_{\rm{1}}&2\delta&\tilde{\Omega}_{\rm{2}}&0&0\\
0&\tilde{\Omega}_{\rm{2}}&0&\tilde{\Omega}_{\rm{3}}&0\\
0&0&\tilde{\Omega}_{\rm{3}}&2\delta&\tilde{\Omega}_{\rm{4}}\\
0&0&0&\tilde{\Omega}_{\rm{4}}&0\\
\end{array}
\right).
\end{equation}
Based on the above analysis, we formulate a five-level M-STIRSAP method. This protocol requires no additional couplings and, in principle, inherits the core advantages of STA.

\begin{figure}
\centering{\includegraphics[width=10cm]{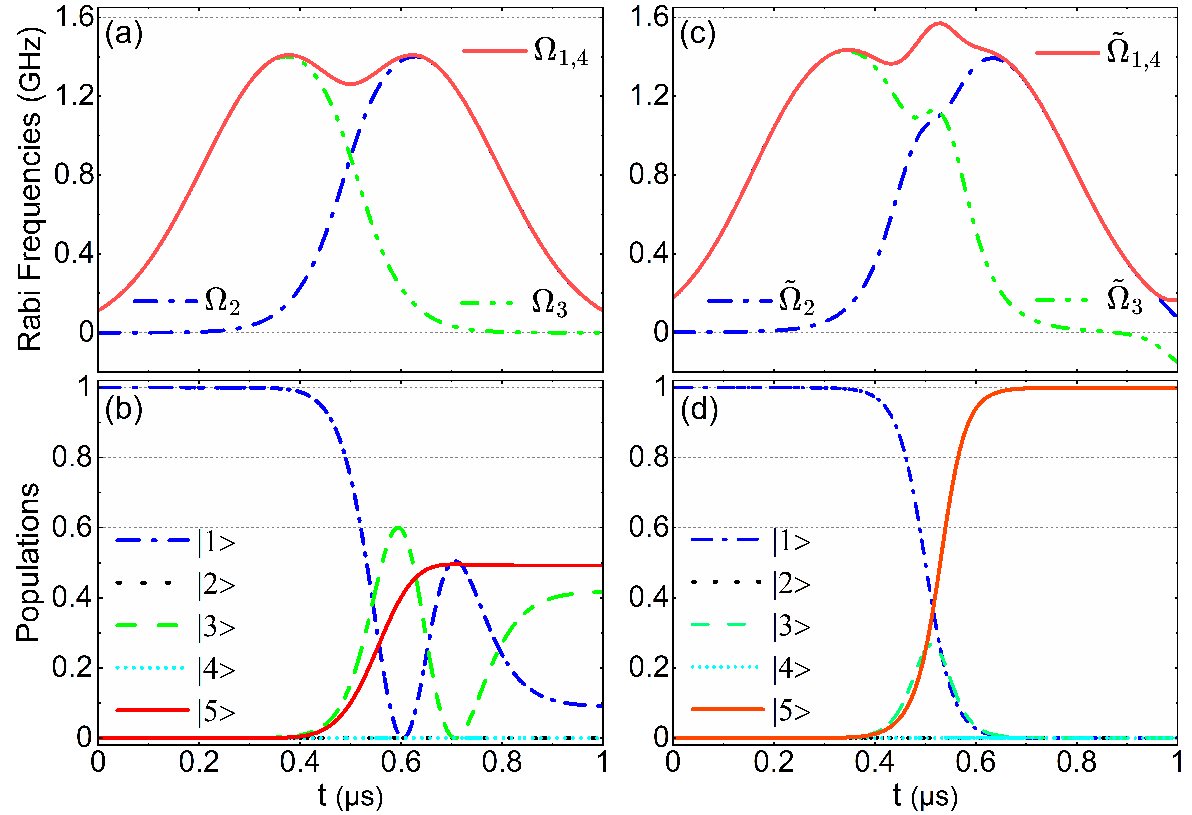}}
\caption{(Color online) Rabi frequencies (upper row) and populations (lower row) as a function of $t$; the left and right columns correspond to M-STIRAP and M-STIRSAP, respectively. Adopted Parameters: $\omega_0\!=\!2\pi\!\times\!5$MHz, $t_f\!=\!1\mu s, \tau\!=\!t_f/8, \sigma\!=\!t_f/6$, $\Delta\!=\!2\mathrm{\pi}\!\times\!5$GHz.}
\label{fig2}
\end{figure}

In order to verify the validity of our protocol, we substitute the Hamiltonians in Eqs.~(\ref{1}) and~(\ref{16}) into the TDSE and numerically study population evolution via the Runge-Kutta $\rm{4}$ algorithm.
%In order to illustrate the validity of our method, we substitute the Hamiltonians in Eqs.~(\ref{1}) and~(\ref{17}) into the TDSE to numerically investigate the population evolution by using Runge-Kutta $\rm{4}$ algorithm.
As a typical example, we assume the Rabi frequencies in Eqs.~(\ref{5}) as
\begin{eqnarray}\label{17}
\omega_{1}\!=\!\omega_{0}\exp\left[-\frac{(t\!-\!t_{f}/2\!-\!\tau)^{2}}{\sigma^{2}}\right],\nonumber\\
\omega_{2}\!=\!\omega_{0}\exp\left[-\frac{(t\!-\!t_{f}/2\!+\!\tau)^{2}}{\sigma^{2}}\right].
\end{eqnarray}
Once they are determined, the original Rabi frequencies in Eq.~(\ref{1}) are obtainable via Eqs.~(\ref{2}) and~(\ref{6}), while the modified ones in Eq.~(\ref{16}) follow from Eqs.~(\ref{15}). %The simulation parameters are set as $t_f\!=\!1\mu s,\omega_0\!=\!2\pi\!\times\!5\mathrm{MHz}, \tau\!=\!t_f/8, \sigma\!=\!t_f/6, \Delta\!=\!2\pi\!\times\!5\mathrm{GHz}$.

Figure.~\ref{fig2}(a) and~(c) show the original Rabi frequencies and the resulting population dynamics for M-STIRAP protocol. As clearly shown in the figure, the dynamical evolution governed by the original Hamiltonian fails to satisfy the adiabatic condition, leading to imperfect population transfer. In contrast, if we utilize four modified Rabi frequencies given by Eqs.~(\ref{15}), a complete population transfer will be realized within a short timescale, as depicted in Figs.~\ref{fig2}(b) and~(d). %In addition, the AE principle ensures that the dynamical evolution of the system is decoupled from the two excited states throughout the process, which effectively suppresses losses arising from these two states.

\begin{figure}
\centering{\includegraphics[width=10cm]{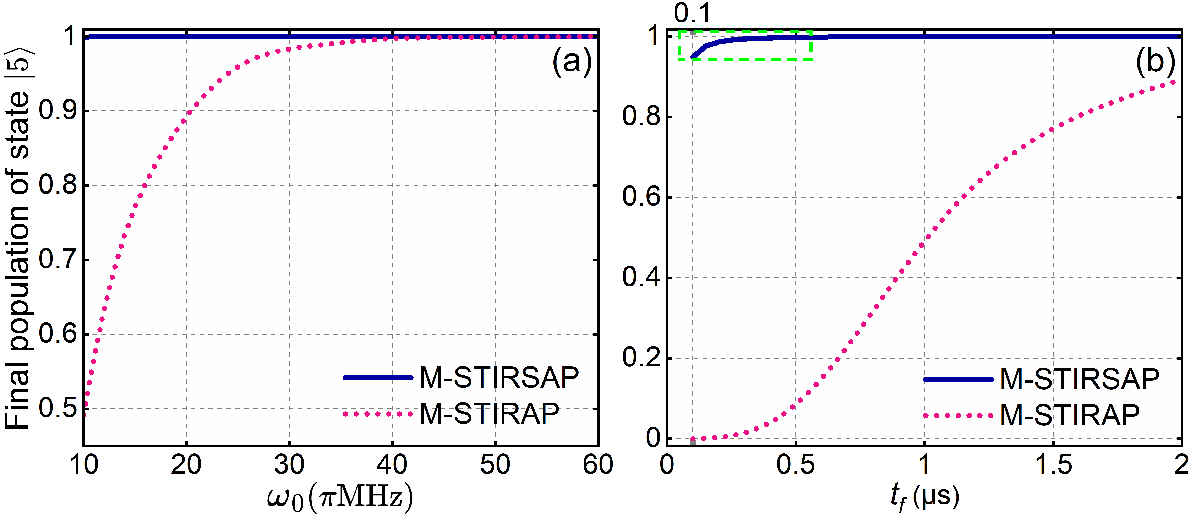}}
\caption{(Color online) The final population of state $|5\rangle$ as functions of parameter (a) $\omega_0$ and (b) $t_f$ under the M-STIRAP ang M-STIRSAP, respectively. All the other parameters are the same as those in Fig.~\ref{fig2}.}
\label{fig3}
\end{figure}

Further studies can reveal more features of the M-STIRSAP protocol.
In Fig.~\ref{fig3}(a), we study the final population of state $|5\rangle$ as a function of $\omega_0$ under M-STIRSAP and M-STIRAP, respectively, while in Fig.~\ref{fig3}(b), we study the final population of state $|5\rangle$ as a function of $t_f$ under these two protocols.
From this comparison, we can find that the performance of M-STIRSAP is always superior to that of M-STIRAP. %To realize high transfer efficiency, M-STIRAP requires a larger pulse area. %To achieve high efficiency, M-STIRAP demands a larger pulse area, i.e., larger average Rabi frequencies or equivalently longer interaction times, which is a well-known limitation of adiabatic nature.

The above results reveal that the M-STIRAP protocol exhibits good performance in terms of robustness and speed. Nevertheless, we find that shortening
time will yields unfavorable outcomes. To this end, we will introduce the TR method in the following section to address this problem.

\subsection{\label{sec:level2.2}The TR-modified M-STIRSAP}
In this section, we integrate the TR method into M-STIRSAP to optimize its driving fields, aiming to realize high-efficiency population transfer within a shorter time. We term this optimized protocol M-STIRSAP-opt.

Let us refocus on the effective three-level Hamiltonian Eq.~(\ref{12}), which evolves over the time interval $t\in[0, t_f]$ and is governed by the TDSE. The corresponding unitary evolution is given by
% the combination of the two ideas has been applied for constructing STAP.
%Clearly, it evolves within the time interval $\tau\in[0, t_f]$ and described by the TDSE. The corresponding unitary evolution is given by:
\begin{eqnarray}\label{18}
\hat{U}(t_f)={\hat{\mathscr{T}}}\exp\left\{-i\int_{0}^{t_f}\tilde{H}_3(t') dt' \right\}.
\end{eqnarray}
where ${\hat{\mathscr{T}}}$ refer to the time-ordering operator. We call the dynamics generated by the Hamiltonian in Eq.~(\ref{12}) the reference process. The full evolution time of this process is given by $\Delta t_{ref}\!=\!t_f$. Next, we introduce a scaling function $t\!=\!f(\mathrm{\tau})$ to reshape the dynamical behavior of the system, which enables a redefinition of the temporal evolution. This transformation allows us to reformulate the evolution operator as
\begin{eqnarray}\label{19}
\hat{\mathscr{U}}\left[f^{-1}(t_f)\right]&={\hat{\mathscr{T}}}\exp\left\{-i\int_{f^{-1}(0)}^{f^{-1}(t_f)}{\tilde{H}_3}\left[f(\tau)\right] \dot{f}(\tau) d\tau \right\}\nonumber\\
&={\hat{\mathscr{T}}}\exp\left\{-i\int_{f^{-1}(0)}^{f^{-1}(t_f)} \bar{H}_3(\tau) d\tau \right\},
\end{eqnarray}
%where the integral limits satisfy $\tau\in\left[f^{-1}(0), f^{-1}(t_f)\right]$, and
where $\bar{H}_3(\tau)\!=\!\tilde{H}_3\left[f(\tau)\right]\dot{f}(\tau)$ denotes the TR Hamiltonian. We refer to the dynamics generated by this Hamiltonian as the TR process, which requires a duration of $\Delta\tau\!=\!f^{-1}(t_f)\!-\!f^{-1}(0)$ to finish.

To ensure the TR process proceeds faster than the reference process, we impose the constraints listed below:
(\romannumeral1) the initial times coincide: $f^{-1}(0)\!=\!0$;
(\romannumeral2) TR process proceeds faster than the reference process: $f^{-1}(t_f)\!<\!t_f$ (equivalently, $\Delta\tau\!<\Delta t_{ref})$;
(\romannumeral3) the initial and final Hamiltonians must be the same: $\bar{H}_3(0)\!=\!\tilde{H}_3(0), \bar{H}_3[f^{-1}(t_f)]\!=\!\tilde{H}_3(t_f)$.

To satisfy all constraints, one can construct a function in the following form~\cite{PhysRevResearch.2.013133}:
\begin{eqnarray}\label{20}
f(t)\!=\!at\!-\!\frac{(a\!-\!1)}{2\pi a}t_f\sin\left(\frac{2\pi a}{t_f}t\right),
\end{eqnarray}
where $a$ denotes the $time$ $contraction$ $parameter$.

\begin{figure}
\centering{\includegraphics[width=10cm]{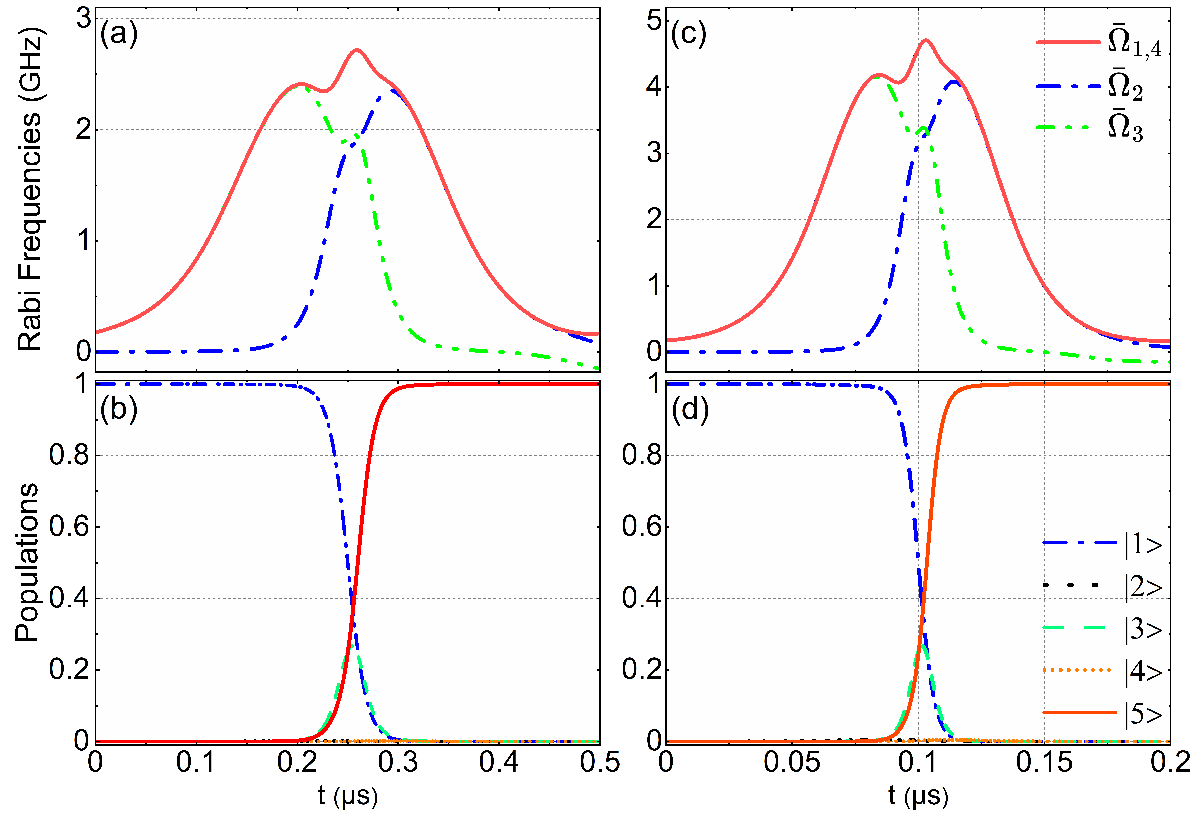}}
\caption{(Color online) Rabi frequencies (upper row) and populations (lower row) as a function of $t$. The left and right columns correspond to
$a\!=\!2$ and $a\!=\!5$, respectively, with all other parameters are the same as those in Fig.~\ref{fig2}.}
\label{fig4}%失谐量在此图中统一了
\end{figure}

Accordingly, the three-level TR Hamiltonian can be written as
\begin{eqnarray}\label{21}
\bar{H}_3\!=\!\frac{\rm{1}}{\rm{2}}
\left(
\begin{array}{ccc}
0&\bar{\omega}_{\rm{1}}&0\\
\bar{\omega}_{\rm{1}}&0&\bar{\omega}_{\rm{2}}\\
0&\bar{\omega}_{\rm{2}}&0\\
\end{array}
\right),
\end{eqnarray}
in which the effective Rabi frequencies become
\begin{eqnarray}\label{22}
&\bar{\omega}_{\rm{1}}\!=\!\tilde{\omega}_{\rm{1}}[f(t)]\dot{f}(t),\quad
&\bar{\omega}_{\rm{2}}\!=\!\tilde{\omega}_{\rm{2}}[f(t)]\dot{f}(t).
\end{eqnarray}

Once again, we are devoted to constructing physically feasible driving fields for the original system. By analogy with Eqs.~(\ref{13})-(\ref{15}), we can directly express them as
\begin{eqnarray}\label{23}
\bar{\Omega}_{1, 4}\!=\!\Big\{4\delta^2\left[\bar{\omega}^2_{\rm{1}}\!+\!\bar{\omega}^2_{\rm{2}}\right]\Big\}^{\frac{1}{4}},\nonumber\\
\bar{\Omega}_{2}\!=\!\bar{\omega}_{1}\left[\frac{4\delta^2}
{\bar{\omega}^2_{1}\!+\!\bar{\omega}^2_{2}}\right]^{\frac{\rm{1}}{\rm{4}}},\nonumber\\
\bar{\Omega}_{3}\!=\!\bar{\omega}_{\rm{2}}\left[\frac{4\delta^2}
{\bar{\omega}^2_{\rm{1}}\!+\!\bar{\omega}^2_{\rm{2}}}\right]^\frac{1}{4}.
\end{eqnarray}

\begin{figure}
\centering{\includegraphics[width=10.5cm]{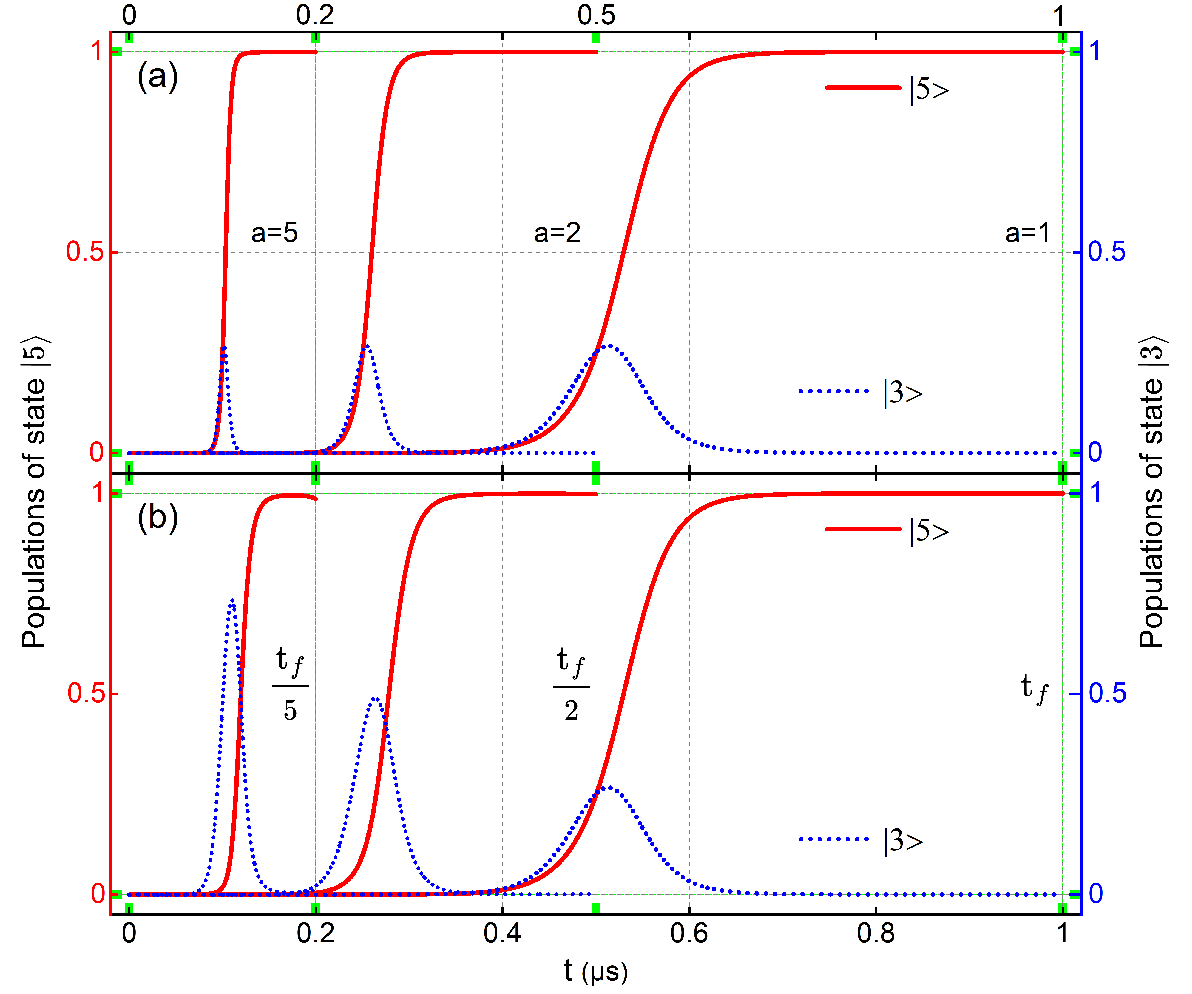}}
\caption{(Color online) (a) Population evolution of the intermediate state $|3\rangle$ and the final state $|5\rangle$ under the M-STIRSAP-opt protocol when $a=1,2,5$; (b) Population evolution of the intermediate state $|3\rangle$ and the final state $|5\rangle$ under the M-STIRSAP protocol for different $t_f$. All other parameters are consistent with those in Fig.~\ref{fig2}.}
\label{fig5}
\end{figure}

With the newly designed Rabi frequencies and the adopted AE condition $\delta\!\gg\!\bar{\Omega}_{j}(j\!=\!1,2,3,4)$, the five-level TR Hamiltonian is written as
\begin{equation}\label{24}
\bar{H}_5\!=\!\frac{\rm{1}}{\rm{2}}
\left(
\begin{array}{ccccc}
0&\bar{\Omega}_{\rm{1}}&0&0&0\\
\bar{\Omega}_{\rm{1}}&2\delta&\bar{\Omega}_{\rm{2}}&0&0\\
0&\bar{\Omega}_{\rm{2}}&0&\bar{\Omega}_{\rm{3}}&0\\
0&0&\bar{\Omega}_{\rm{3}}&2\delta&\bar{\Omega}_{\rm{4}}\\
0&0&0&\bar{\Omega}_{\rm{4}}&0\\
\end{array}
\right).
\end{equation}

In Fig.~\ref{fig4}, we take $a\!=\!2$ and $a\!=\!5$ as typical examples to verify the validity of the M-STIRSAP-opt protocol. It is obvious that complete population transfer can be realized in both cases, and their dynamical evolutions are faster than that of the reference process shown in Fig.~\ref{fig2}(d).

To better demonstrate the advantage of the TR-based strategy, we compare the dynamical behaviors of the M-STIRSAP and M-STIRSAP-opt protocols under the same evolution time. %To ensure a fair comparison, three different values of parameter $a$ ($a\!=\!1,2,5$) are considered in the upper panel, while three cases with $t\!=\!t_f\!=\!1\ \mathrm{\mu s}$, $t\!=\!t_f/2\!=\!0.5\ \mathrm{\mu s}$, and $t\!=\!t_f/5\!=\!0.2\ \mathrm{\mu s}$ are considered respectively in the lower panel.
Figures.~\ref{fig5}~(a) and~(b) present the temporal population dynamics of states $|3\rangle$ and $|5\rangle$ for the M-STIRSAP-opt and M-STIRSAP protocols, respectively.
We observe that only when $a=1$(corresponding to $t\!=\!t_f\!=\!1\mathrm{\mu s}$) are the two protocols equivalent and thus show the same population dynamics. As the evolution time shortens, however, the gap between them grows far more noticeable. These differences are mainly reflected in two aspects: the final population of state $|5\rangle$ and the transient population of state $|3\rangle$. In contrast, M-STIRSAP-opt works very well and, more intriguingly, exhibits population time behaviors highly similar to that of the reference protocol within a shorter timescale.

\begin{figure}
\centering{\includegraphics[width=12cm]{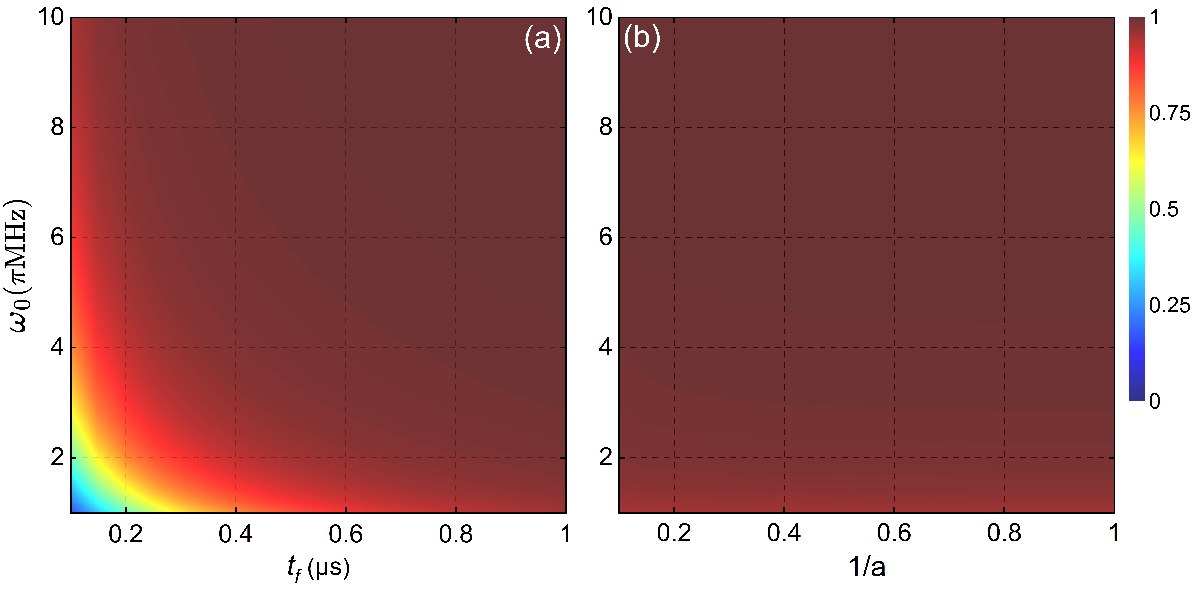}}
\caption{(Color online) (a) Contour plot of the final population of state $|5\rangle$ as a function of $t_f$ and $\omega_0$ for M-STIRSAP; (b) Contour plot of the final population of state $|5\rangle$ as a function of $t_f$ and $1/a$ for M-STIRSAP-opt. All other parameters are the same as those in Fig.~\ref{fig2}.}
\label{fig6}
\end{figure}

Next, we further evaluate the robustness of both protocols.
Figure.~\ref{fig6}(a) provides a contour plot of the final population of state $|5\rangle$ as a function of $t_f$ and $\omega_0$ for the M-STIRSAP protocol. The results show that ideal population transfer is realized only when the parameter values are close to those adopted in Fig.~\ref {fig2}; any deviation from the ideal parameter values will degrade the control performance.
As a comparison, Fig.~\ref{fig6}(b) presents a contour plot of the final population of state $|5\rangle$ as a function of $1/a$ and $\omega_0$ for the M-STIRSAP-opt protocol. The results clearly reveal that nearly perfect population transfer efficiency can always be obtained against parameter fluctuation.
Therefore, the M-STIRSAP-opt holds promising potential to maintain favorable control performance and outperforms the M-STIRSAP protocol.

\begin{figure}
\centering{\includegraphics[width=10cm]{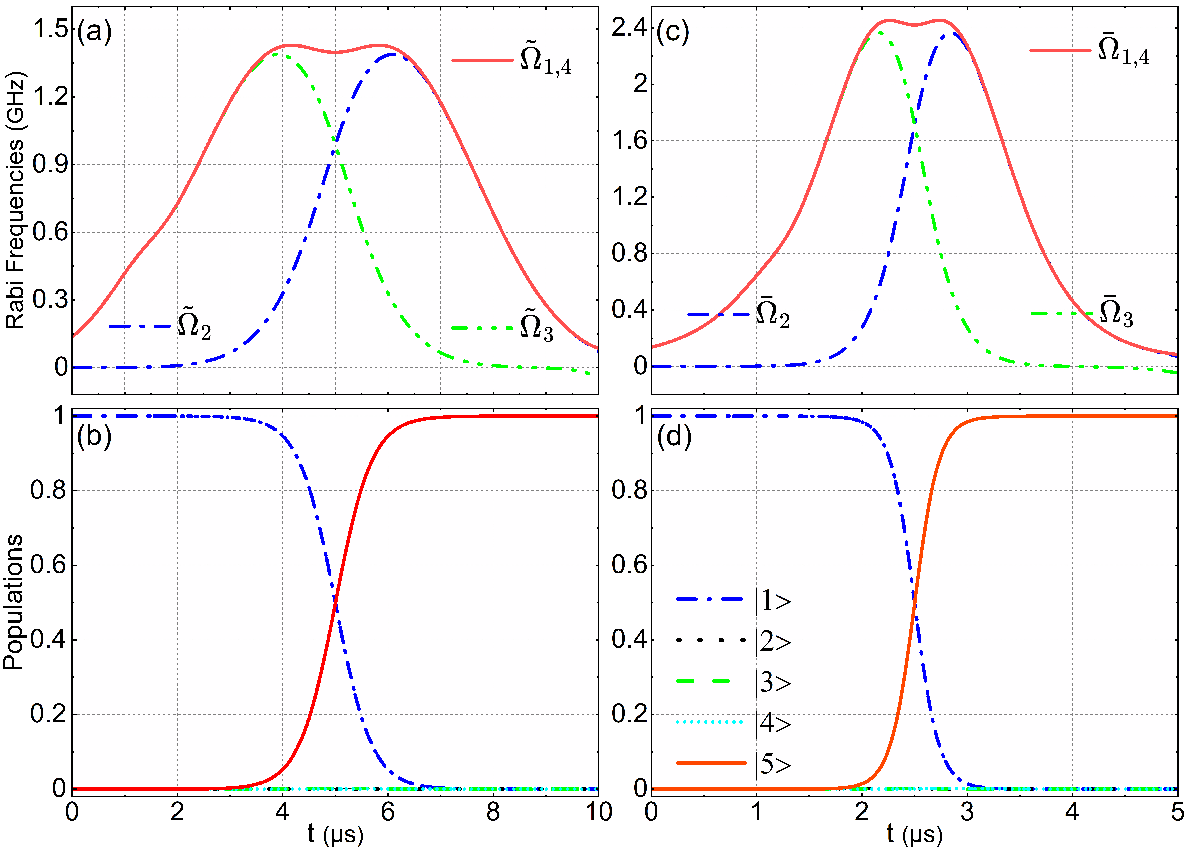}}
\caption{(Color online) Rabi frequencies (upper row) and populations (lower row) as a function of $t$; the left and right columns correspond to M-STIRSAP and M-STIRSAP-opt, respectively. Adopted Parameters: $\omega_0\!=\!2\pi\!\times\!5$MHz, $t_f\!=\!10\mu s, \tau\!=\!t_f/10, \sigma\!=\!t_f/6$, $\Delta\!=\!2\mathrm{\pi}\!\times\!5$GHz.}
\label{fig7}
\end{figure}

Once fast and robust population transfer from $|1\rangle$ to $|5\rangle$ is achieved, we therefore set out to suppress transient populations of the intermediate states over the full dynamical evolution. Notably, the AE protocol inherently decouples the system dynamics from the two loss-prone excited levels, leaving only the third level as our key target for population suppression. Herein, we propose to first implement the above objective via the M-STIRSAP protocol, followed by speeding up the evolution process using the TR-optimized M-STIRSAP-opt protocol. The corresponding results are clearly presented in Fig.~\ref{fig7}, from which we can find that all intermediate levels are almost completely unpopulated.  This phenomenon benefits the realization of efficient and robust population transfer between the initial and final states of a chainwise-connected five-level system. Throughout the entire population transfer procedure, the utilized pulses exhibit smooth and singularity-free profiles, ensuring favorable experimental realizability~\cite{PhysRevA.100.043413, Zheng2022}. Furthermore, the proposed theoretical scheme requires a much shorter implementation timescale compared with the experimentally realized three-level STIRSAP process~\cite{Du2016}, which provides room for reasonable trade-offs in practical tests.

\section{Summary}
In summary, we have developed two multi-state stimulated Raman shortcut-to-adiabatic passage protocols for efficient and fast population transfer in chain-connected five-state systems. To construct these protocols without introducing extra couplings, we first reduce the original five-level system into equivalent three- and two-level counterparts. On this basis, the first protocol is implemented by combining CD driving and unitary transformation, while the second protocol adopts the time-rescaling method to provide an optimized solution. Our results show that both protocols work well, while the second can achieve superior performance within a shorter timescale compared with the first protocol. Furthermore, both protocols are able to reduce transient populations on all intermediate levels. As such, we believe this work can serve as a valuable supplement to existing methods for the coherent control of chainwise-connected five-level systems.

%may benefit applications built upon multi-state quantum systems.

\section{Acknowledgments}
We would like to thank the anonymous referees for constructive comments that are helpful for improving the quality of the work. This work was supported by the National Natural Science Foundation of China (Grants No.~12475026, No.~12075193, and No.~12365004) and the Doctoral Research Startup Fund of Northwest Normal University (Grant No.~6014/202503101301).

\section{Data availability statement}
All data that support the findings of this study are included
within the article (and any supplementary files).

\bibliographystyle{unsrt}
\bibliography{references}
\end{document}